# Overcoming I/O bottleneck in superconducting quantum computing: multiplexed qubit control with ultra-low-power, base-temperature cryo-CMOS multiplexer

R. Acharya[1,2], S. Brebels[1], A. Grill[1], J. Verjauw[1,3], Ts. Ivanov[1], D. Perez Lozano[1], D. Wan[1], J. Van Damme[1,2], A. M. Vadiraj[1], M. Mongillo[1], B. Govoreanu[1], J. Craninckx[1], I. P. Radu[1], K. de Greve[1,2], G. Gielen[1,2], F. Catthoor[1,2], A. Potočnik*[1]

[1]Imec, Kapeldreef 75, Leuven, B-3001, Belgium
[2]Department of Electrical Engineering (ESAT), KU Leuven, Leuven, B-3000, Belgium
[3]Department of Materials Engineering (MTM), KU Leuven, Leuven, B-3000, Belgium

## Abstract

Large-scale superconducting quantum computing systems entail high-fidelity control and readout of large numbers of qubits at millikelvin temperatures, resulting in a massive input-output bottleneck. Cryo-electronics, based on complementary metal-oxide-semiconductor (CMOS) technology, may offer a scalable and versatile solution to overcome this bottleneck. However, detrimental effects due to cross-coupling between the electronic and thermal noise generated during cryo-electronics operation and the qubits need to be avoided. Here we present an ultra-low power radio-frequency (RF) multiplexing cryo-electronics solution operating below 15 mK that allows for control and interfacing of superconducting qubits with minimal cross-coupling. We benchmark its performance by interfacing it with a superconducting qubit and observe that the qubit's relaxation times ($T_1$) are unaffected, while the coherence times ($T_2$) are only minimally affected in both static and dynamic operation. Using the multiplexer, single qubit gate fidelities above 99.9%, i.e., well above the threshold for surface-code based quantum error-correction, can be achieved with appropriate thermal filtering. In addition, we demonstrate the capability of time-division-multiplexed qubit control by dynamically windowing calibrated qubit control pulses. Our results show that cryo-CMOS multiplexers could be used to significantly reduce the wiring resources for large-scale qubit device characterization, large-scale quantum processor control and quantum error correction protocols.

## Introduction

Quantum information processing on error-corrected quantum computers promises a significant computational advantage over classical computers in solving particular problems that are otherwise intractable[1–4]. Although many physical systems are under active investigation[5–7], superconducting quantum circuits have emerged as arguably the most developed platform[8–11] with recent proof-of-principle demonstration of verifiable quantum advantage[12–14]. State-of-the-art superconducting quantum processors[15,16] use external circuits with at least one control line per qubit running from the room-temperature stage to the base-temperature stage of a dilution refrigerator (Fig. 1a). This approach can be extended up to 100 qubits[16] but is hardly scalable to build a large-scale quantum computer that requires dynamic circuit operations, such as quantum error correction and active reset on a large number of high-fidelity qubits[17–19], due to the limited number of input-output ports on the dilution refrigerator[20]. Additionally, such an approach will also limit the throughput for large-scale statistical characterization of quantum devices at millikelvin temperatures, which will be crucial to enable qubit fabrication at scale.

An architecture involving multiplexing elements employed at the base-temperature stage of a dilution refrigerator can alleviate such an input-output bottleneck for qubit characterization and control[20–23] (Fig. 1b). Such functional multiplexing has already been demonstrated for solid-state spin-qubit systems, where quantum devices can also be co-integrated with multiplexers[24–29] and control electronics[30–32], albeit at elevated temperatures (1-4K) to account for the finite heat dissipation of the electronics. However, stringent requirements of superconducting quantum processors in terms of low electromagnetic and thermal noise[15,16] makes such direct co-integration with dissipative cryo-electronics, such as multiplexers, extremely challenging[21,22]. Consequently, a practical multiplexing solution for superconducting quantum devices must present cryogenic compatibility at operating temperatures below 50 mK, with ultra-low noise and low-power consumption, while ensuring large operational bandwidth, high dynamic range, and fast switching for multiplexed control at microwave frequencies. To that end, several technologies have been explored, such as mechanical switches[33], superconducting nanowires[34] and Josephson-junction based switches[35,36]. Promising results were demonstrated with two-dimensional electron-gas based RF switching using capacitive depletion of the electron gas or high-electron-mobility-transistor (HEMT)[37]. However, the low on-off ratio of ~20 dB in the 4-8 GHz bandwidth of interest rendered them unsuitable for superconducting qubits. Encouraging developments have also been made in paradigm shifting approaches using optical links[38] or rapid single flux quantum (RSFQ) electronics[39]. In spite of this, it is unclear if these technologies can mature enough to meet the demands of a large-scale superconducting quantum computing system.

The cryo-CMOS platform, on-the-other-hand, offers a versatile, scalable and mature solution that is readily available to be leveraged in quantum computing[21–23,40]. RF signal multiplexing at millikelvin temperatures has also been demonstrated using cryo-CMOS based devices[41–43]. However, until now, their performance was not tailored to meet the stringent requirements of superconducting qubits.

Here we demonstrate a cryo-CMOS based multiplexing solution operating at the base temperature of a dilution refrigerator. Our solution combines the requisite bandwidth, speed, dynamic range and noise performance for functional interfacing with superconducting qubits, while minimally loading and heating the fridge. In particular, we report a DC-to-10 GHz single-pole-4-throw (SP4T) RF multiplexer based on 28-nm bulk CMOS technology and extensively verify its operation for qubit control and interfacing. The multiplexer is operated at the base-temperature stage of a dilution refrigerator, and both its ultra-low static (0.6 µW) and dynamic (~0.48 pJ/Hz of switching) power dissipation result in minimal heating of the system of only a few mK, despite the minute power handling capabilities of the fridge at these extremely low temperatures. In terms of RF behavior, we observe both a high isolation (35 dB at 6 GHz) and a low insertion loss (3 dB at 6 GHz), confirming its viability as a multiplexer for superconducting qubits. We verify the interfacing capability with superconducting qubits by measuring a qubit with coherence times exceeding 20 µs and single qubit gate fidelities surpassing 99.9%. In addition, and importantly, we demonstrate that active time-division multiplexing of qubits is possible using our multiplexer, and verify its performance by selectively windowing a calibrated qubit pulse with over 30 dB port-to-port crosstalk suppression.

## Design and characterization of ultra-low-power cryo-CMOS multiplexer

The multiplexer chip is custom designed and fabricated in a commercial foundry using 28 nm bulk CMOS process (see Methods). The multiplexer has four RF input ports (RF1-4) and a common RF output port (RFC) that can be programmed using either serial or parallel control, as schematically depicted in Fig. 1c. The roles of the ports can also be reversed to operate the device as a de-multiplexer. The micrograph of the multiplexer chip, along with the different building blocks of the multiplexer and the physical location of the ports are shown in Fig. 1 d,e.

In contrast to the previous design[41], a high degree of isolation between the RF ports is provided by employing two levels of series-shunt switches, as illustrated by the simplified circuit schematic in Fig. 2a. The isolation is further enhanced by minimizing the mutual coupling between the printed circuit board (PCB) tracks and the bondwires. This is achieved using resonant LC structures made of the parasitic pad capacitance and on-chip inductance, with resonant frequency of 6 GHz. The number of electrostatic discharge (ESD) protection cells is reduced to four to provide the requisite protection against ESD events while significantly reducing power consumption[41]. The static bias resistance (>1 kΩ) at the gate of the RF switch transistors, required to avoid RF signal leakage, is shunted by means of a pair of diode-connected transistors for faster switching times (see Supplementary Section 1 and Supplementary Fig. 1).

The multiplexer PCB is anchored and thermalized to the 10 mK base-temperature stage of a dilution refrigerator (see Methods and Extended Data Fig. 1 for details about the measurement setup). The static power consumption of the multiplexer is measured as a function of the applied bias voltage $V_{dd}$ and plotted in Fig. 2b. The multiplexer turns on at a threshold voltage of ~0.6 V and has an ultra-low power consumption of 0.60 µW at $V_{dd}$ = 0.7 V (red circles). This dissipation results in a slight increase in the fridge temperature of less than 2 mK. Approximately 60% of the temperature increase can be attributed to the dissipation in the ESD protection circuit (gray diamonds, also see Supplementary Section 2 and Supplementary Fig. 2), which is not directly related to the size of the multiplexer and would therefore allow for scaling to larger systems.

Dynamic switching between the RF ports, required for fast control or readout multiplexing, results in an additional power dissipation due to the charging and discharging of the internal capacitive nodes in the circuit. Fig. 2c shows the dynamic power consumption (with static offset removed) measured as a function of the multiplexer switching rate for $V_{dd}$ of 0.7 V and 0.9 V. An extremely low dynamic power consumption of ~1pW/Hz/V$^2$ (~0.48 pJ/Hz at Vdd = 0.7 V) is obtained, which is dominated by the gate capacitance of the bulky RF switch transistors (see Supplementary Section 3 and Supplementary Fig. 3). Whereas in static or low-frequency systems such as spin qubits[27,37], cryo-electronic multiplexing solutions were previously demonstrated with little to no effect on the base temperature, extending these results to the dynamic high-frequency regime of superconducting qubits requires different circuit designs, accounting for the novelty of our approach.

Next, the multiplexer's RF performance is characterized by measuring the transmission $S_{21}$ between the ports RFC and RF3 (see Methods). The port-to-port isolation is measured as a function of the signal frequency and plotted in Fig. 2d. The isolation shows a weak frequency dependence with a minimum value of 30 dB and maximum value exceeding 40 dB at 0.7 V, most likely due to reflections and impedance mismatches in the cryogenic measurement setup. The inset of the plot shows that the normal operation begins around the threshold voltage of 0.6 V and the port-to-port isolation is approximately constant above 0.7 V. To obtain the insertion loss through the port RF3, a reference measurement is performed by disconnecting the multiplexer from the signal path in a subsequent cooldown. An average insertion loss of 2.3 dB is obtained in the 4-8 GHz bandwidth as shown in Fig. 2e. Expectedly, it exhibits a dependence on the bias voltage similar to the port-to-port isolation (see inset of Fig. 2e).

These results show that both the dynamic and static power dissipation have a strong dependence on operating voltage, whereas the RF performance of the multiplexer is practically constant above the threshold voltage. Hence, a significant reduction in the power consumption can be achieved by designing devices for lower threshold voltages without compromising the RF performance. For example, by extrapolating the measurement data to $V_{dd}$ = 0.3 V, the estimated static and dynamic power dissipation would be 30 nW and 90 fJ/Hz, respectively. In this regard, a process technology with lower or tunable threshold voltage, such as fully-depleted-silicon-on-insulator (FDSOI), offers an attractive platform for the realization of future devices and is being actively investigated for use in millikelvin electronics[28,44]. Our results also show that design optimizations, in particular, the ESD protection cells and the RF switching transistors, can further reduce the power dissipation. Additionally, employing advanced process nodes with shorter channel lengths and lower channel resistance could potentially enhance performance.

## Cryo-CMOS multiplexer thermal noise benchmarking using high-coherence qubits

The effective carrier temperature of cryo-CMOS devices operating at millikelvin temperatures is, in general, higher than the lattice temperature[28,45]. Therefore, introducing these devices for delivering control signals to superconducting qubits can adversely affect the qubit performance. It disrupts the low noise environment designed for qubits in state-of-the-art dilution refrigerator systems[16]. To quantify these effects, the noise performance of the multiplexer is benchmarked using a fixed-frequency high-coherence superconducting transmon qubit ($\omega_q/2\pi$ = 3.957 GHz), fabricated using a recently reported fab-compatible superconducting qubit fabrication process[46]. The standard quantum circuit quantum electrodynamics (cQED) architecture is adopted[47], where the qubit is dispersively coupled to a readout resonator ($\omega_r/2\pi$ = 6.471 GHz), which in turn is coupled to a feedline for control and readout (see Supplementary Fig. 4). The relevant device parameters are listed in Extended Data Table. 1. All the required RF signals for qubit manipulation and readout are sent via the feedline. The multiplexer is connected to the feedline via a directional coupler and a total of 13 dB of attenuation to reduce the emanating noise seen by the qubit (see Fig. 3a; also see Methods and Extended Fig. 1 for measurement setup). The dominant effect of the noise associated with the higher effective carrier temperature in the multiplexer is that it results in (noisy) blackbody radiation that is partially absorbed by the attenuator and then injected into the resonator – where it results in an increase of the photon number fluctuations i.e., photon shot noise in the readout resonator. This, in turn, results in an enhanced dephasing of the qubit due to the ac-Stark effect[15,48]. To establish a baseline and to enable standalone characterization of this noise, a direct RF path that bypasses the multiplexer is used (see Fig. 3a). In this configuration, the qubit acts as a sensitive noise detector and the multiplexer behaves as a voltage and switching-frequency dependent noise source.

The dependence of the qubit's energy relaxation time ($T_1$), Ramsey decoherence time ($T_2^*$) and spin-echo ($T_2^e$) decoherence time is measured as a function of the bias voltage $V_{dd}$ of the multiplexer. These characteristic lifetimes show no dependence on $V_{dd}$ until the multiplexer turns on at the threshold voltage of ~0.6 V as shown in Fig. 3b. The $T_1$ remains unaffected even above this voltage, maintaining a mean value of ~30 µs, while the $T_2^*$ and $T_2^e$ show a notable voltage dependence and decrease from ~40 µs to ~25 µs at the nominal operating voltage of $V_{dd}$ = 0.7 V. From the decoherence rate $\Gamma$ (=$1/T_2^e$), the effective noise temperature[48] of the multiplexer of ~150 mK and a thermal photon flux of ~0.15 photons/√Hz at the frequency of the readout resonator is extracted (see Methods). The effect of this electronic heating on the qubit performance can be further minimized in future applications by additional attenuation. For example, a mere 20 dB of total attenuation between the multiplexer and the qubit would push the $T_2$ decoherence limit set by the multiplexer to over 400 µs – much longer than the intrinsic decoherence channels. In addition, it is also verified that pair-breaking quasiparticle generation[49] mechanisms are not at play (see Supplementary Fig. 5), which is also indicated by the independence of $T_1$ as a function of $V_{dd}$.

The measurement results obtained for $T_1$, $T_2^*$ and $T_2^e$ on characterizing the qubit with signals routed through the multiplexer operating at $V_{dd} = 0.7$ V is shown in Fig. 3d. Even in this configuration, the obtained qubit lifetimes are similar to the values measured using the direct line (Fig. 3d). This result is very encouraging as it establishes that the multiplexer noise can be treated independently, thus validating the adopted approach of characterizing the noise by measuring the qubit using the direct RF line and using the multiplexer exclusively as a noise source.

The described characterization approach is especially useful to assess the effect of the multiplexer noise generated by dynamically switching between its ports. The multiplexer is programmed to operate in the parallel operation mode (see Supplementary Section 3) and the control line D0 is continuously switched so that the multiplexer switches between the signal paths through ports RF1 and RF2. The obtained $T_1$, $T_2^*$ and $T_2^e$ values of the qubit as a function of the multiplexer switching rate is shown in Fig. 3e. Remarkably, the $T_1$ times still show no apparent dependence up to the measured switching rate of 1 MHz. On the other hand, the $T_2^*$ and $T_2^e$ times decrease from the static-mode value of ~25 μs to ~10 μs at 1 MHz of switching, corresponding to an effective multiplexer temperature of ~500 mK and a thermal photon flux of ~1.10 photons/√Hz at the frequency of the readout resonator. Nevertheless, using a total attenuation of 20 dB after the multiplexer, the decoherence limit set by the multiplexer could be pushed to over 50 μs.

In future experimental scenarios involving multi-qubit devices[11], the multiplexer is advocated to be used for either multiplexed readout or multiplexed control of superconducting qubits. In the former case, the multiplexer will only be switched during the readout pulses at a rate dictated by the readout pulse duration, which for a fast high-fidelity readout is on the order of a MHz[11,50]. For the latter case, the multiplexer will have to switch much faster at a maximum rate defined by the typical qubit gate lengths of 10-50 ns. However, it is crucial to note that in this configuration, the multiplexer will drive the qubit through the dedicated X-Y control line[15] (charge-line) and not via the feedline as demonstrated above. The multiplexer noise in the X-Y control line couples transversely to the qubit quantization axis[15], thereby affecting its $T_1$ and not $T_2$. For an average switching rate of 20 MHz and standard charge-line coupling parameters (see Methods), a limit of ~50 μs on $T_1$ can be estimated without attenuation, which could further be increased to ~5 ms with an attenuation of 20 dB, far beyond the state-of-the-art energy relaxation times[9].

## Cryo-CMOS multiplexer for high-fidelity time-multiplexed quantum gates

The CMOS multiplexer is intended for quantum computing applications that demand high-fidelity static and dynamic qubit operations. To assess the impact of the multiplexer on single qubit gate fidelities, randomized benchmarking[51,52] (RB) is performed on the qubit under various operating conditions of the multiplexer, that encompass both static biasing and dynamic switching. For each operating condition, the RB gate fidelity is measured using gate pulses of duration $t_g = 40$ ns and plotted as a function of qubit's $T_2^*$ in Fig. 4a. The inset of the plot shows the individual curves obtained with the multiplexer turned off (green) and multiplexer switching at 5 MHz with $V_{dd} = 0.7$ V (violet). The data clearly shows two distinct regimes: for low $T_2^*$ (<10 μs) gate fidelities are limited by the coherence time of the qubit with an inversely proportional dependence to the obtained coherence time (black dashed line), while for long $T_2^*$ the fidelity saturates at ~99.93% (gray dashed line) due to the imperfect pulse calibration[53,54]. Nevertheless, for nominal operating conditions of the multiplexer with dynamic switching of 1 MHz, $T_2^*$ exceeds 10 μs and the measured gate fidelity is above 99.9%. The limit set by coherence time can be further increased with enhanced thermal filtering of the multiplexer noise and longer qubit lifetimes, whereas the threshold set by the pulse calibration errors can be improved by using shorter control pulses, adjusting the pulse shapes, or using different Clifford gate sets[55,56].

In large-scale quantum systems, each qubit invariably requires a dedicated control line for state manipulation. The resulting hardware cost and wiring complexity for large systems can be significantly reduced by multiplexing the qubit control lines. However, this introduces an additional source of crosstalk as the drive signals intended for one qubit could leak through the multiplexing network to another qubit. This needs to be avoided for high-fidelity qubit operations. Additionally, operations involving more than one qubit will require time-division multiplexing of the control signals to the appropriate qubits. The corresponding output port needs to be active for an optimal time window around the control pulse to perform the intended qubit operation. Windows larger than the pulse width would result in wasted dead-time between the pulses and windows shorter than the pulse width would distort it, reducing the gate fidelity in both cases.

To demonstrate the capability to implement time-division multiplexing and to quantify the potential limitations mentioned above, a novel experiment with a pulse sequence shown in the top panel of Fig. 4b is performed. The

RF signals are exclusively routed through the multiplexer, which is initially programmed to port RF2 (D0 = 1). A fixed pre-calibrated 40 ns cosine-enveloped π-pulse drives the qubit from ground state to excited state, when routed entirely to the qubit through port RF1 (D0 = 0). The amount of pulse excitation seen by the qubit is modified by reducing the signal time window centered around the π-pulse by dynamically switching the port D0 of the multiplexer.

To quantify the crosstalk, the excited state population $p_e$ of the qubit is measured and plotted as a function of the multiplexer time-window (pink, left y-axis) in the bottom panel of Fig. 4b. The data shows excellent agreement with numerical simulations performed in Qutip[57], where instantaneous switching of the multiplexer with a rise/fall time of 0 ns is assumed (solid black line). In the absence of switching (i.e., window time of 0 ns), the obtained excited state population is below the detection threshold of the setup, indicating that the qubit remains in the ground state. This strongly establishes that time-multiplexed qubit control can indeed be performed using the multiplexer. Even when ascribing the entire measured qubit population to crosstalk, a lower bound of 30 dB on port-to-port isolation is obtained, which agrees well with the reported values (see Fig. 2d). The deviation of the excited state population from simulations at short time windows can be attributed to the finite rise and fall times of the multiplexer of ~2.6 ns, resulting in a smaller excited state population (see Supplementary Fig. 1b,c for rise time measurements). It is worthwhile to note that in practical devices, qubits are generally not degenerate but are detuned with different transition frequencies[11,12]. This translates to easing the crosstalk constraint as the qubit is insensitive to (crosstalk) signals that are not close to its own transition frequency.

To quantify the minimum time-window required for proper operation, the data is further explored by analyzing $1 - p_e$ as a function of the multiplexer time window in Fig. 4b (turquoise, right y-axis). A noticeable decrease in the measured excited state population of the qubit is observed for multiplexer windows below 30 ns, albeit with a small offset to numerical simulations owing to the finite rise and fall times of the multiplexer. For time windows over 30 ns, there is no discernable difference in the measured excited state population of the qubit due to the shallow pulse tails of the cosine envelope. Therefore, the length of the multiplexer switching window does not have to exceed the qubit control pulse width. This result is substantial as it enables pulse sequencing without any inactive time between qubit control pulses to account for multiplexer switching delays.

It should be noted that time-division multiplexing of qubit control or readout inherently introduces an additional overhead in pulse sequencing. However, this is not completely detrimental as physical circuits implementing quantum algorithms[11,12] do not require constant access to all the qubits or the readout resonators. A recent circuit compilation study[58] reported that it is possible to minimize this potential overhead, if not completely eliminate it.

Now that the compatibility of the cryo-CMOS multiplexer with superconducting qubits has been established, it is imperative to survey its scalability to larger systems. Commercially available dilution refrigerators have cooling powers in the order of 20 µW at the base-temperature stage. The presented cryo-CMOS multiplexer dissipates around 0.24 µW of static power, barring the contribution of ESD cells, and 0.5 µW of dynamic power with 1 MHz switching at 0.7 V. Therefore, assuming ~0.2 µW of power consumption per qubit channel, it would be possible to interface it with approximately 100 qubits within the cooling budget of the refrigerator. With a reduction in threshold voltage and the consequent reduction in operating voltage to 0.3 V, the power consumption per output channel reduces to ~25 nW, increasing the interfacing capability to almost 1000 qubits. Scaling this number to a million qubits with current cooling powers is still extremely demanding, with power dissipation limited to ~20 pW per qubit channel. Therefore, further investigation into design and operation optimization, along with research into advanced process technologies with tunable threshold voltage, is required to understand the physical limitations on the extent of scalability.

## Conclusions

We have demonstrated and extensively characterized an ultra-low power cryo-CMOS multiplexer fabricated using commercially available 28 nm bulk CMOS technology that operates at the base-temperature stage of a dilution refrigerator. The remarkable RF and noise properties of the multiplexer over a wide-frequency range enables unprecedented interfacing with high-coherence superconducting qubits to perform qubit control with single-qubit gate fidelities exceeding 99.9%. The nanosecond-scale fast-switching of the multiplexer enables time-division multiplexing with signal crosstalk suppressed by greater than 30 dB. Our results pave the way for a viable path to address the wiring bottleneck in future large-scale quantum computing systems.

## Methods

## Chip design

The cryo-CMOS multiplexer is designed by performing room temperature and -40°C device simulations, the lowest temperature at which the transistor models are reliable. Chip-level circuit simulations are done using Cadence Virtuoso. Electromagnetic simulations are performed using Integrated EMX for on-chip passives.

## Measurement setup

The cryo-CMOS multiplexer and the qubit are measured with a Bluefors LD400 dilution-refrigerator setup (see Extended Data Fig. 1). Two separate input lines are used for qubit measurements and each of them is thermalized using 20-dB attenuators at two different temperature stages. High frequency noise above the measurement frequency range is filtered before reaching the qubit with a low-pass filter (VLF-8400+) with cut-off frequency of 8.4 GHz. One input line is directly connected to the qubit via the -20-dB port of a directional coupler (Krytar 120420). The other input line is routed to the qubit via the multiplexer, a 10-dB attenuator, the directional coupler, and an additional 3-dB attenuator. The attenuators are thermalized to the base plate of the fridge using copper braids. The multiplexer ports D0 and D1 are programmed in parallel operation mode with low-pass filtered (VLFX-780+; cut-off frequency 780 MHz) AWG (Keysight M3202A) signals. Output signal lines are thermalized with two isolators (LNF-ISC4_8A) and a circulator (LNF-CIC4_8A) with a total reverse isolation of ~ 60 dB and a 4-8 GHz band-pass filter (KBF-4/8-2S). The signal is amplified with a HEMT amplifier (LNF-LNC4_8C) at the 4 K stage and an ultra-low-noise amplifier (LNA-30-04000800-07-10P) at room temperature. Pulsed measurement signals are generated and acquired using the Keysight Quantum Engineering Toolbox: M3202A AWGs, M3102A digitizer and M9347A Dual DDS local oscillators). The qubit excitation and readout pulse are combined at room temperature and applied to the qubit's feedline. No dedicated charge line or flux line are used to excite or bias the qubits.

The superconducting qubit chip is wire-bonded to a non-magnetic gold-plated copper PCB enclosed in an oxygen-free copper sample holder. The sample holder is thermalized to the mixing chamber plate in a dilution refrigerator and surrounded by a copper radiation shield as well as two cryo-perm shields to minimize the magnetic fields at the sample. The cryo-CMOS multiplexer chip is placed on a four-layer printed circuit board (PCB) with low-loss Rogers RO4350B microwave laminate.

## Cryo-CMOS multiplexer: measurement and analysis

The power dissipation of the multiplexer is determined by measuring the voltage drop across a 1 kΩ series resistor on the power supply line of the multiplexer using Keithley 2182A nanovoltmeter. The series current inferred using Ohm's law is used to calculate the power dissipation.

The RF characterization of the multiplexer is limited to 4-8 GHz due to the bandpass filter and the high-electron-mobility-transistor (HEMT) amplifier present in the output line. The port-to-port isolation is obtained by measuring the transmission $S_{21}$ between the ports RFC and RF3 while sequentially programming the multiplexer from port RF1 through to port RF4.

## Thermal photon number and effective carrier temperature estimation

The thermal photon number $n$ in the readout resonator is estimated using the resonator line-width $\kappa_r$, the dispersive shift[15] $\chi$, and the dephasing rate $\Gamma$ as [48]:

$$n = \Gamma \frac{\kappa_r}{4\chi^2} \frac{\kappa_r^2 + 4\chi^2}{\kappa_r^2}. \tag{1}$$

The various parameters and their values are listed in Extended Data Table. 1. The thermal photon number is translated to effective carrier temperature of the multiplexer using the Bose-Einstein distribution and the signal attenuation. Similarly, following the inverse procedure, the effective temperature can be translated to dephasing rate.

## $T_1$ limit due to multiplexer thermal noise

Using the drive-line coupling[59] $A_d$ to the qubit and the voltage noise power spectral density $S_{VV}(\omega)$[60] at the qubit frequency, the $T_1$ is estimated as[61]:

$$T_1 = \frac{1}{\Gamma_1} = \frac{\hbar^2}{A_d^2 S_{VV}(\omega_q)}. \tag{2}$$

Where
$$S_{VV}(\omega) = \frac{4R_m \hbar \omega}{e^{\frac{\hbar \omega}{k_B T_{eff}}} - 1} \quad (3)$$

and
$$A_d = \sqrt{\frac{\hbar C_q \omega_q}{2}} \frac{C_d}{C_d + C_q}. \quad (4)$$

Using typical qubit-chargeline coupling capacitance $C_d$ of 0.1 fF, qubit capacitance $C_q$ of 110 fF, multiplexer resistance $R_m$ of 5 Ω and an effective multiplexer carrier temperature of 7 K at 20 MHz switching with $V_{dd}$= 0.7 V, a limit on the relaxation time of 50 μs is obtained.

## Randomized benchmarking: measurement and analysis

The average physical gate error $r_g$ and gate fidelity $F_{1q}$ is measured by applying a random sequence of Clifford gates of varying lengths $n$ to the qubit initialized in the ground state. At the end of the sequence, an inverting gate is added to create an overall identity operation and the final qubit state is measured to determine the fidelity. The measurement is repeated 80 times for each sequence length. Cosine pulses with DRAG pulse calibration[56] and a duration of 40 ns are used in the experiment. The averaged sequence fidelity is fitted to $F = Ap^n + B$, where the parameters $A$ and $B$ depend on the state preparation and measurement errors[62]. From $p$, $r_g$ is determined as the error per Clifford $r_{Clifford}$ normalized by the average number of physical Clifford generator gate (1.875):

$$r_g = \frac{r_{Clifford}}{1.875} = \frac{(1-p)(d-1)}{1.875\,d}. \quad (5)$$

Where $d = 2^m$ is the dimensionality of the Hilbert space, which is equal to 2 for a single qubit. The average physical gate fidelity is obtained from:

$$F_{1q} = 1 - r_g. \quad (6)$$

Fidelity as a function of the gate duration $t_g$, relaxation time $T_1$ and Ramsey decoherence time $T_2^*$ is modelled using[53,63]:

$$F_{1q-fit} = 1 - c_0 - \frac{k_1}{T_\phi^{mux}}. \quad (7)$$

Where
$$\frac{1}{T_\phi^{mux}} = \frac{1}{T_2^*} - \frac{1}{T_2^*|_{V_{dd}=0\,V}}. \quad (8)$$

$c_0$ accounts for the gate error due to relaxation ($= t_g/3T_1$) and other dephasing channels, $T_\phi^{mux}$ is the dephasing time due to the multiplexer, $T_2^*$ is the measured decoherence time and $k_1$ a scaling parameter that depends on the noise power spectral density (PSD) seen by the qubit. The obtained value for fitting parameter $k_1$ is approximately $0.433 t_g/3$. For a white noise PSD, the noise correlation time is small compared to the experimental timescales and the factor $k_1$ reduces to $t_g/3$. However, photon shot noise has a Lorentzian PSD[48,60] with two-photon correlation time is in the order of a microsecond (inverse of resonator linewidth $1/\kappa_r$), which being comparable with experimental timescales results in a deviation from the white noise case.

## Acknowledgements
The authors gratefully thank the imec P-line, operational support, and the MCA team. This work was supported in part by the imec Industrial Affiliation Program on Quantum Computing.

## Author's contribution
R.A. and A.P. planned the experiment. S.B. designed the multiplexer. A.P. designed the qubit samples. Ts.I, D.P.L. fabricated the qubit samples, with contributions from D.W. R.A. and A.P. performed the measurement and analysis of qubit data at base-temperature. S.B. and A.G. performed the measurements and analysis of the ESD protection cells from room-temperature to 4 K. R.A., A.P., J.V., J.V.D. and A.M.V. prepared the experimental setup and methods. R.A. and A.P. prepared the manuscript, with input from all authors. A.P., I.R., J.C., K.dG., B.G., M.M., G.G. and F.C. supervised and coordinated the project.

## Data availability
The data that support the plots within this paper and other finding of this study are available from the corresponding authors upon reasonable request

# Figures

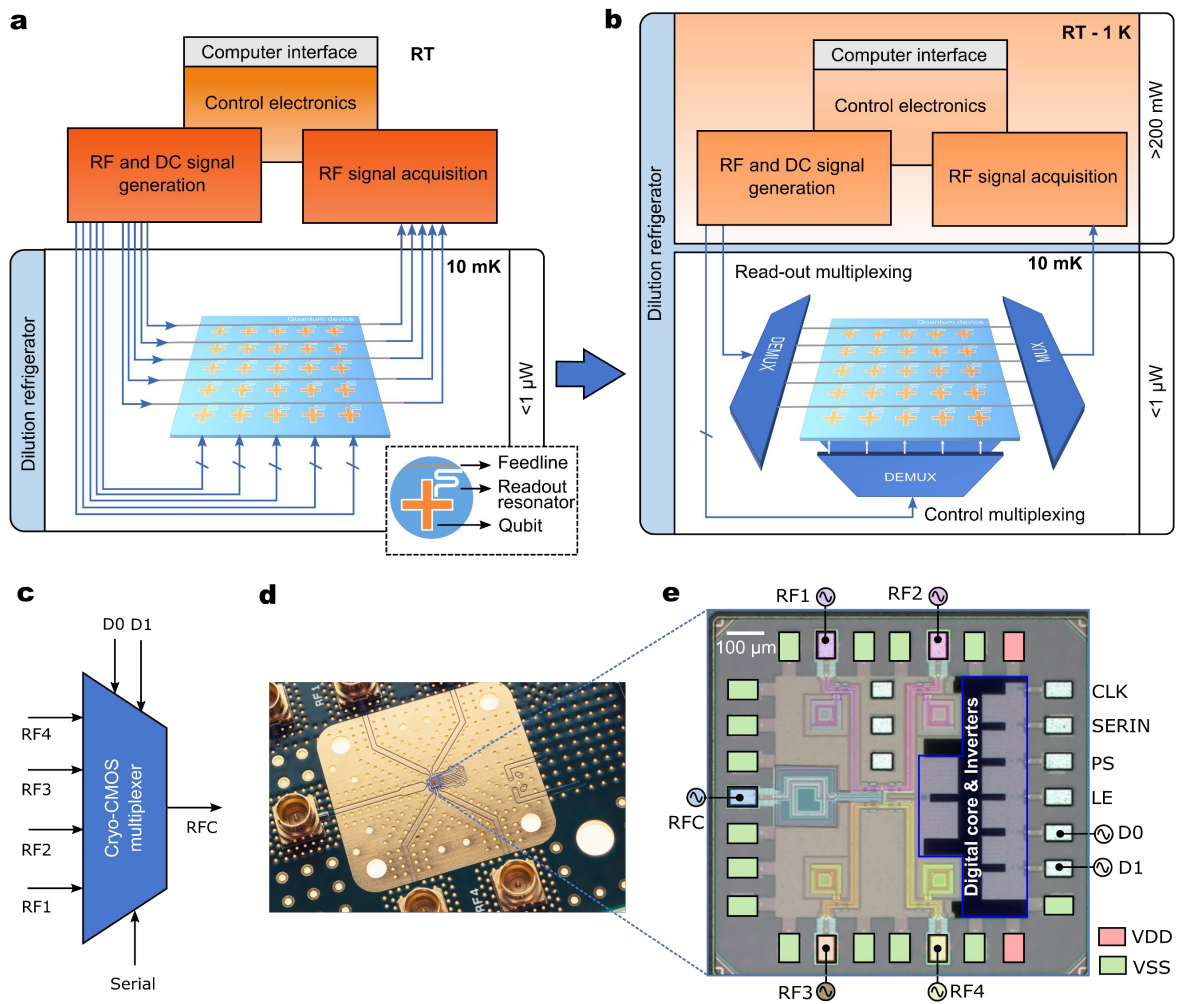

**Fig. 1| Routing microwave signals using cryo-multiplexers. a,** Standard RF signal routing for measuring superconducting qubits in a dilution refrigerator. **b,** Scheme for multiplexing the control and readout signals at the base-temperature stage of a superconducting quantum computer. The required RF signals can be generated from either room-temperature electronics outside the dilution refrigerator or cryo-electronics operating inside it. **c,** Schematic representation of the cryo-CMOS multiplexer. **d,** Optical image of the PCB onto which the cryo-CMOS multiplexer is wirebonded. **e,** Optical micrograph of the cryo-CMOS multiplexer chip. The ports of the multiplexer and the physical location of the digital blocks are labelled. False colors are used in the RF portion of the chip to identify constituent components of the ports RFC and RF1-4.

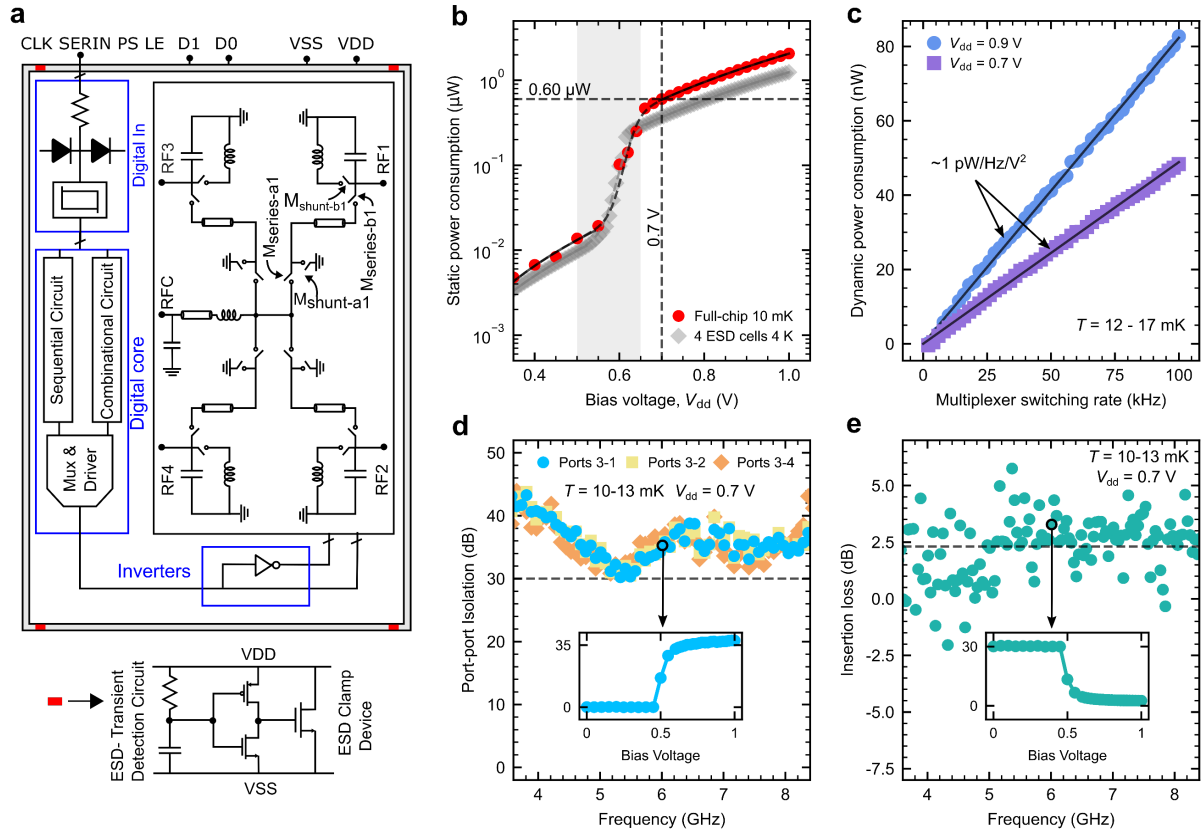

**Fig. 2| Power and RF performance of the cryo-CMOS multiplexer. a**, Simplified circuit schematic of the multiplexer (top panel) and ESD protection cell (bottom panel). The multiplexer can be controlled using a serial interface (CLK, SERIN, PS, LE) or a parallel interface (D1, D0, LE). Two levels of series-shunt switches along with LC resonant circuits using the pad capacitance and an on-chip inductance provide high isolation between RF ports. The minimum number of 4 ESD cells is used to provide sufficient ESD protection while minimizing static power consumption. **b**, Static power consumption as a function of the bias voltage $V_{dd}$ of the multiplexer chip measured at base temperature (red circles) and of the ESD protection cells measured at 4 K (gray diamonds). The gray zone represents the transition region from off to on state with a threshold voltage of the device around 0.6 V. The solid black lines are cubic fits to measurement data. **c**, Dynamic power consumption of the multiplexer versus the switching rates at $V_{dd}$ = 0.9 V (blue) and $V_{dd}$ = 0.7 V (purple). Black solid lines are linear fits to data from which a dynamic power dissipation of 1 pJ/Hz/V$^2$ is extracted. **d**, Port-to-port isolation of the multiplexer measured between port RF3 and other ports (see Methods) versus the frequency at $V_{dd}$ = 0.7 V. The minimum isolation is 30 dB with values approaching 40 dB within the 4-8 GHz bandwidth of interest. The inset shows the isolation measured as a function of bias voltage at 6 GHz. **e**, Insertion loss versus frequency measured at port RF3 at $V_{dd}$ = 0.7 V showing a mean of 2.3 dB (dashed black line) in the 4-8 GHz bandwidth. The inset shows the insertion loss obtained versus bias voltage at 6 GHz.

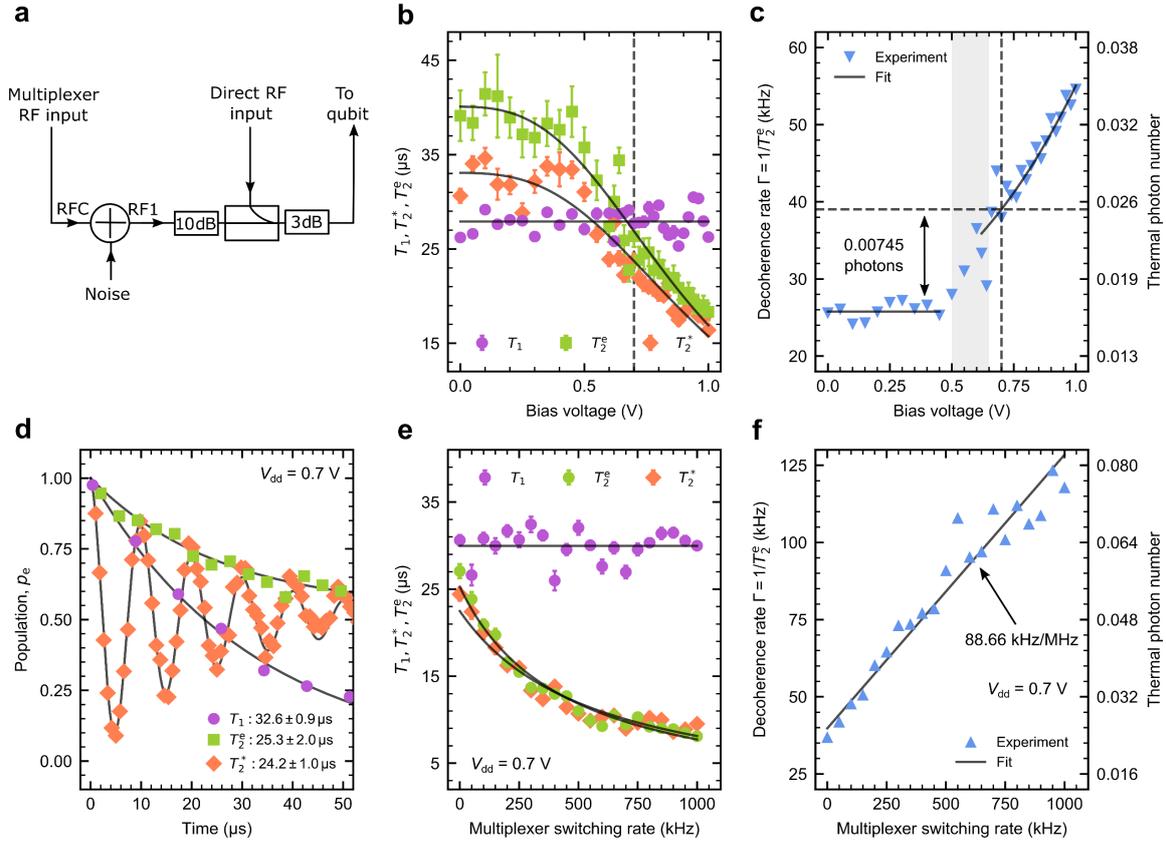

**Fig. 3| Benchmarking the cryo-CMOS multiplexer performance using a high-coherence qubit. a**, Simplified RF signal path (see Extended Fig. 1 and Methods). The qubit can either be measured using the direct RF input line or using the RF line routed through the multiplexer port RF1. The multiplexer also adds a voltage and switching rate dependent noise. **b**, $T_1$, $T_2^*$, $T_2^e$ measured using the direct line as a function of the multiplexer bias voltage. $T_1$ is unaffected but $T_2^*$ and $T_2^e$ decrease from ~35 μs to ~25 μs at $V_{dd}$ = 0.7 V. **c**, Decoherence rate $\Gamma = 1/T_2^e$ versus the bias voltage and the extracted thermal photon numbers at the readout resonator. **d**, $T_1$, $T_2^*$, $T_2^e$ times measured using RF signals routed through the multiplexer at $V_{dd}$ = 0.7 V. The obtained values are similar to the values measured using the direct RF line, confirming that the noise added by the multiplexer is independent of the chosen RF line for qubit measurements. **e**, $T_1$, $T_2^*$, $T_2^e$ values measured versus the switching rate of the multiplexer at $V_{dd}$ = 0.7 V. The $T_1$ is still unaffected while the $T_2$ times further decrease from ~25 μs to ~10 μs at 1 MHz switching. **f**, Extracted decoherence rate and thermal photon number versus switching rate, showing an additional 88.66 kHz of dephasing per MHz of multiplexer switching.

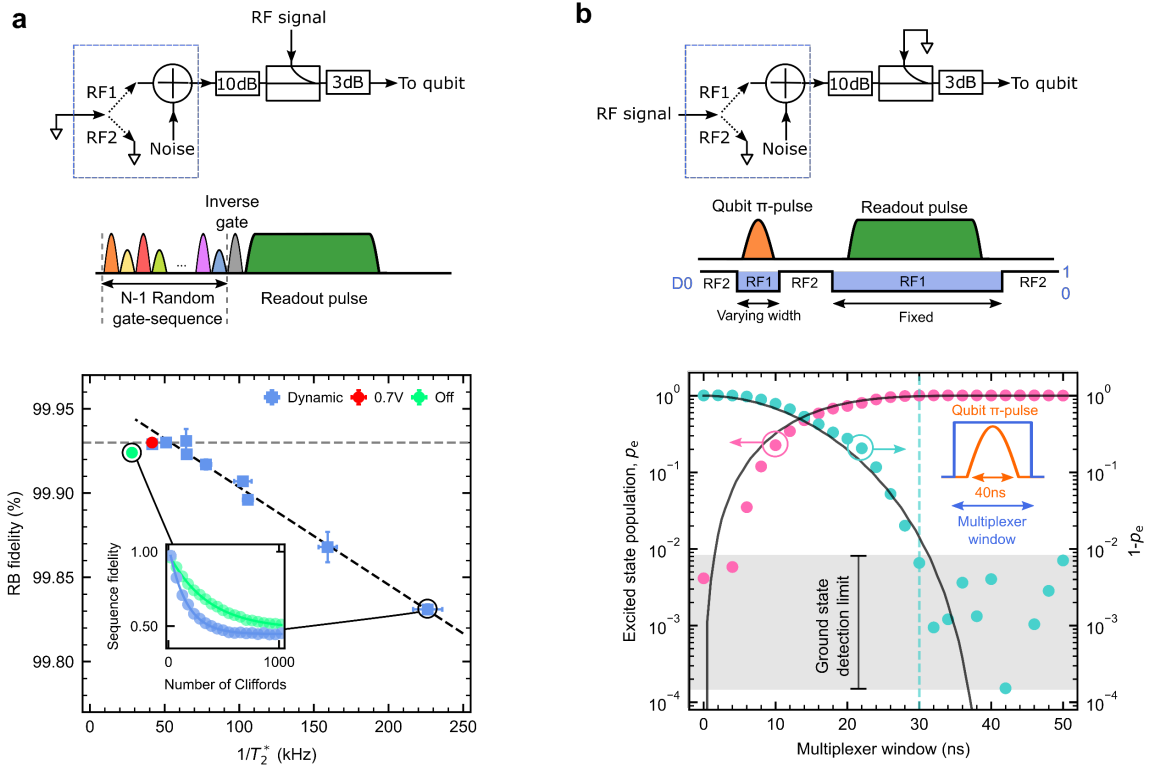

**Fig. 4| Randomized benchmarking and time-domain-multiplexing using cryo-CMOS multiplexer. a**, Simplified signal path and pulse sequence for randomized benchmarking (see Extended Fig. 1 and Methods for more details). Randomized benchmarking (RB) fidelity versus $1/T_2^*$ for the multiplexer OFF (green), ON at 0.7 V (red), ON and switching (blue). The dynamic switching rates from left to right are 55 kHz, 135 kHz, 200 kHz, 265 kHz, 465 kHz, 785 kHz, 1 MHz, 2 MHz and 5 MHz, chosen to get a distribution in decoherence times $T_2^*$. The dashed gray line represents a guide to the eye for the limit in fidelity arising due to finite pulse calibration errors (~ 99.93 %) and dashed black line represents a linear fit to the data as a function of $1/T_2^*$ (see Methods). The inset also shows the individual RB fidelity measurements obtained with a sequence length of 1000 Clifford gates. **b**, Simplified signal path and pulse sequence for the demonstration of time-division multiplexing. The RF signal is normally routed to port RF2 (D0 = 1), except for the time windows in which D0 is set low, in which case it is routed to the qubit via port RF1. Measured excited state population of the qubit, $p_e$, versus the multiplexer window length (pink) shows that below 2 ns time window, the population is below the detection limit of the qubit ground state. The solid black lines are simulation data obtained using a 0 ns rise and fall time for the signal D0. The discrepancy between simulations and experiments at low window times can be explained by the finite rise and fall times, which results in an effective qubit excitation slightly smaller than expected. The plot of $1-p_e$ (turquoise) shows that for time windows above 30 ns, there is no discernable change difference in excited state population, implying that the multiplexer time-window can be exactly as long as the qubit π-pulse.

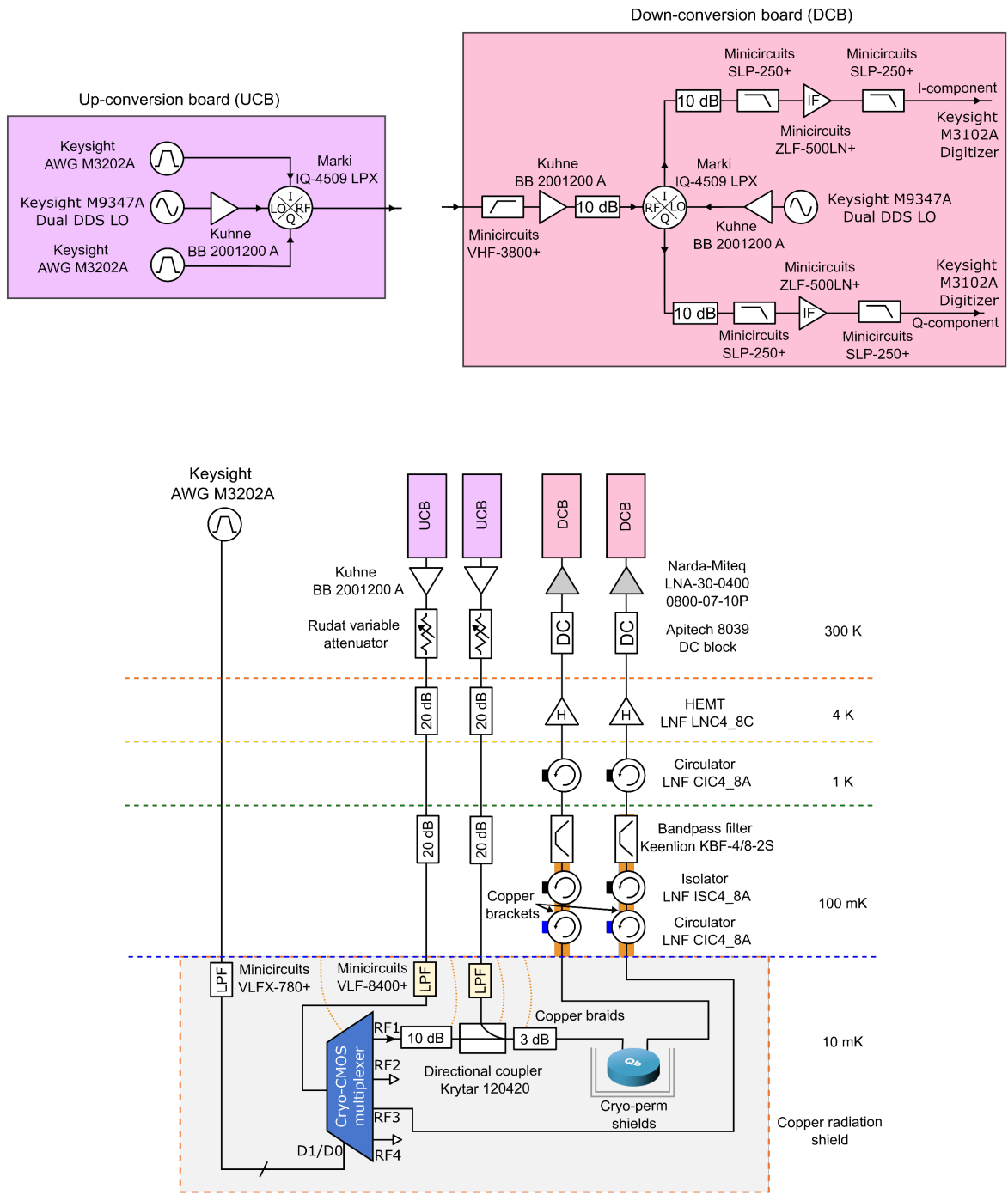

**Extended Data Fig. 1 | Experimental measurement setup.** Room-temperature electronics and dilution refrigerator wiring for the measurement of the cryo-CMOS RF multiplexer performance with superconducting qubits.

**Extended Data Table. 1 | Device parameters**

| Parameter | Notation | Value |
|---|---|---|
| Resonator frequency | $\omega_r/2\pi$ | 6.471 GHz |
| Resonator linewidth | $\kappa_r/2\pi$ | 0.697 MHz |
| Qubit frequency | $\omega_q/2\pi$ | 3.957 GHz |
| Qubit anharmonicity | $\alpha/2\pi$ | -180 MHz |
| Qubit-resonator coupling | $g/2\pi$ | ~ 90 MHz |
| Dispersive shift | $\chi/2\pi$ | -0.259 MHz |

# Supplementary Information

## 1. Optimization and measurement of the multiplexer rise time

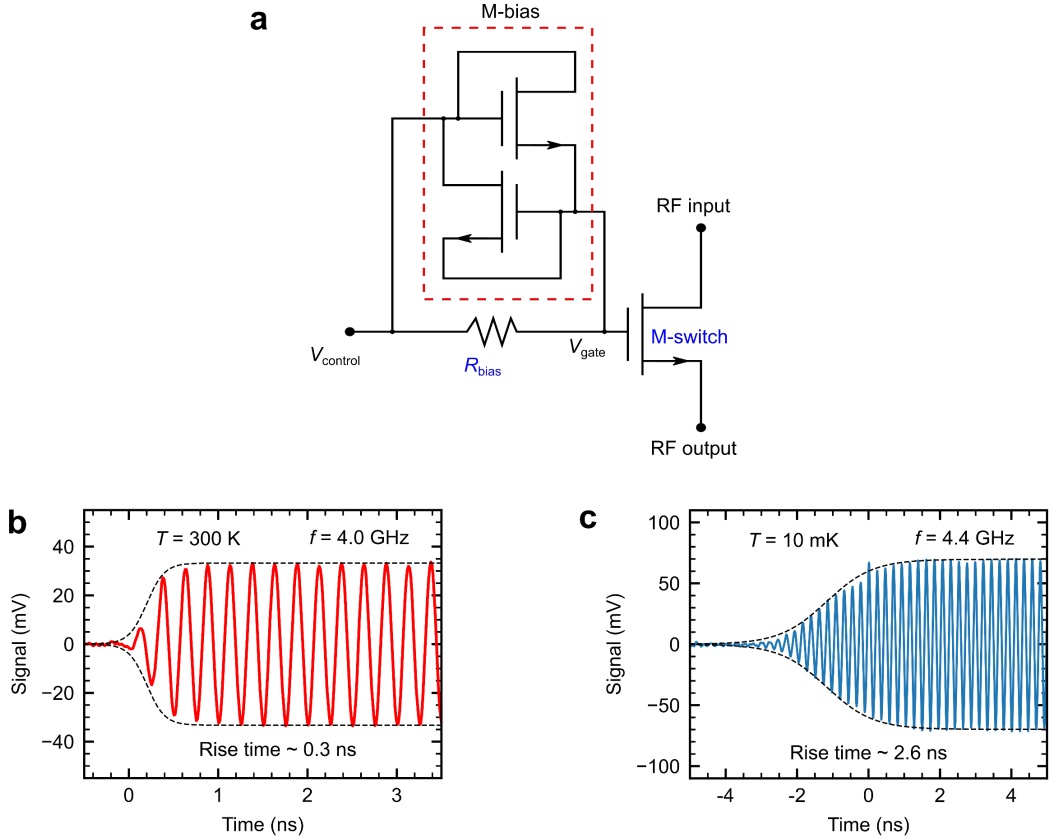

**Supplementary Fig. 1 | Risetime of the cryo-CMOS multiplexer. a**, Nonlinear biasing network using a pair of diode-connected transistors. The bias resistor $R_{bias}$ is shunted by a low-resistance path while $|V_{control} - V_{gate}|$ is greater than the threshold voltage of the shunting network for faster charging and discharging the gate of the switching transistor. **b**, Rise-time measurement of the multiplexer at room temperature. **c**, Rise-time measurement of the multiplexer at the base temperature inside the dilution refrigerator.

A large bias resistor is required at the gate of the RF switch transistor to avoid the leakage of RF signals. This results in a longer switching time due to the large RC time delay. To enable faster switching, the resistor is shunted by a pair of diode-connected transistors as shown in Supplementary Fig. 1a. When $|V_{control} - V_{gate}|$ is greater than the threshold voltage of the diode-connected transistors, a low resistance path is created to charge or discharge the gate of the RF transistor up to the threshold voltage. The required high resistance is restored as soon as $|V_{control} - V_{gate}|$ is less than threshold voltage.

A rise time of ~0.4 ns is measured for the multiplexer at room temperature using an oscilloscope (Supplementary Fig. 1b). This corresponds to the expected rise time due to the delays from the input-output ports, digital logic, and RC time delays. However, the measured rise time inside the dilution refrigerator at 10 mK is ~2.6 ns. This increase is partly attributed to the impedance mismatches in the setup and to the 780 MHz low-pass filter on the control line D0 (See Extended Fig.1).

## 2. Measurement of the ESD leakage power

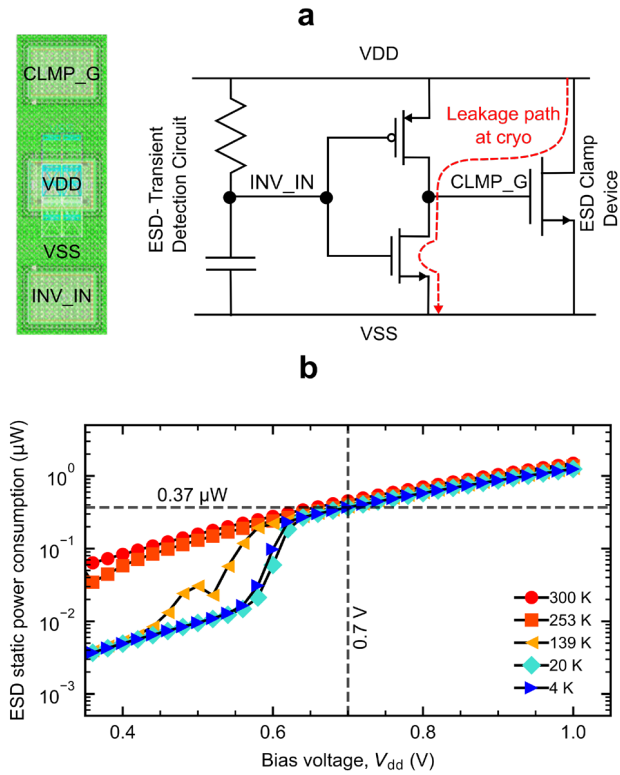

**Supplementary Fig. 2 | Static power dissipation of ESD protection cells. a,** Test chip for measuring the power dissipation of the ESD cells (left) and the RC-based ESD clamp circuit schematic (right). The power-dissipation measured can be attributed to the gate-drain leakage of the ESD clamp transistor. **b,** Static power consumption (scaled to 4 ESD cells) versus the bias voltage measured from room temperature to 4 K using the test-chip. At $V_{dd}$ = 0.7 V, the obtained static power dissipation is 0.37 µW, accounting for more than 60% of the total static dissipation of the full cryo-CMOS multiplexer chip.

The ESD cells are required to protect the core devices during a potential ESD event on the power supply during fabrication, handling or operation. The leakage from the ESD protection circuit is measured using a test chip containing ESD clamp cells from room temperature to cryogenic temperatures up to 4 K (Supplementary Fig. 2a). The test chip has access to the internal node voltages at the input of the inverter and the clamp diode. The obtained static power consumption for a total of 4 ESD protections cells at $V_{dd}$ of 0.7 V is 0.37 µW, accounting for 60% of static power dissipation measured for the full chip at base temperature (Supplementary Fig. 2b). This leakage is attributed to the source-gate leakage of the ESD clamp transistor and verified using simulations (not shown here).

3. Multiplexer programming for serial/parallel mode and dynamic power consumption analysis

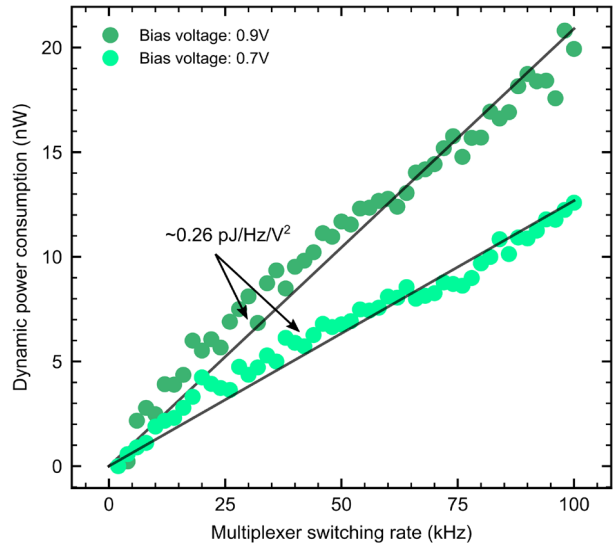

**Supplementary Fig. 3 | Effect of dynamic switching in serial operation mode.**

The choice between serial or parallel programming of the multiplexer can be made using the state of the digital control line PS: high corresponds to serial mode and low corresponds to parallel mode. Serial-mode programming is done using the clock (CLK), serial input (SERIN) and latch enable (LE). After setting PS high, the serial data are moved into an on-chip shift-register, where it is temporarily stored using CLK and SERIN. These data are then latched onto a set of flip-flops on the rising edge of a LE pulse, which then activates the drivers for the proper activation of the RF switches in the serial mode. For parallel operation mode, PS is made low, which in turn resets all the sequential logic. In contrast to serial mode, only combinational logic in used to choose between the RF ports using the control signals D0 and D1, enabling on-demand fast switching. The drivers for the RF switches change their state based on the rate at which switching is performed on ports D0/D1. The combinational gates are not clocked and therefore only have RC delay. The parallel mode requires LE to be set constantly high and can be used as a chip select.

The dynamic power consumption versus the switching rate in parallel operation mode is shown in the main text in Fig. 2c. However, it is not straightforward to disentangle this dissipation into its constitutional digital and RF components. To get a crude estimate, the multiplexer is dynamically switched using D0/D1 while it is programmed in serial operation mode with PS set low (see Supplementary Fig. 3), such that the states of the RF switches and its drivers are not changed. The difference in the obtained dynamic power dissipation in serial-mode (~0.26 pJ/Hz/V$^2$) and parallel-mode (~1 pJ/Hz/V$^2$) suggests that approximately 75 % of the power corresponds to charging and discharging the bulky gate of the RF transistors.

## 4. Qubit sample

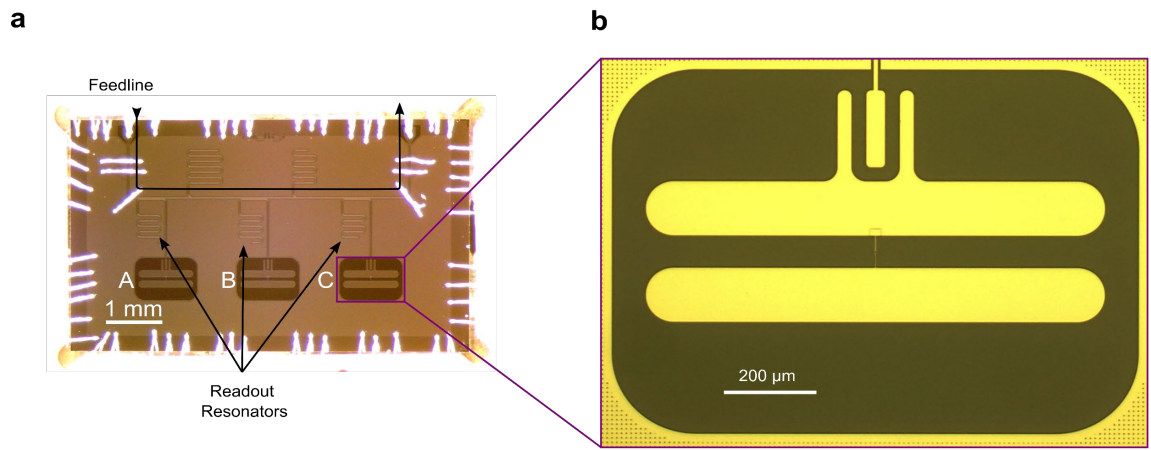

**Supplementary Fig. 4 | Superconducting transmon qubit sample. a,** Optical micrograph of the wire-bonded chip. It consists of three double-pad superconducting transmon qubits labelled A-C made of aluminum, each capacitively coupled to its own readout resonator. All resonators are in turn coupled to a common feedline through which the measurements are performed. The experimental results in the main text correspond to qubit B. **b,** Zoom-in optical micrograph of an individual transmon. The two superconducting islands for the capacitive pads are shunted by a Josephson junction. More details on the device fabrication can be found in Ref. [1]

## 5. Detection of quasiparticle tunnelling

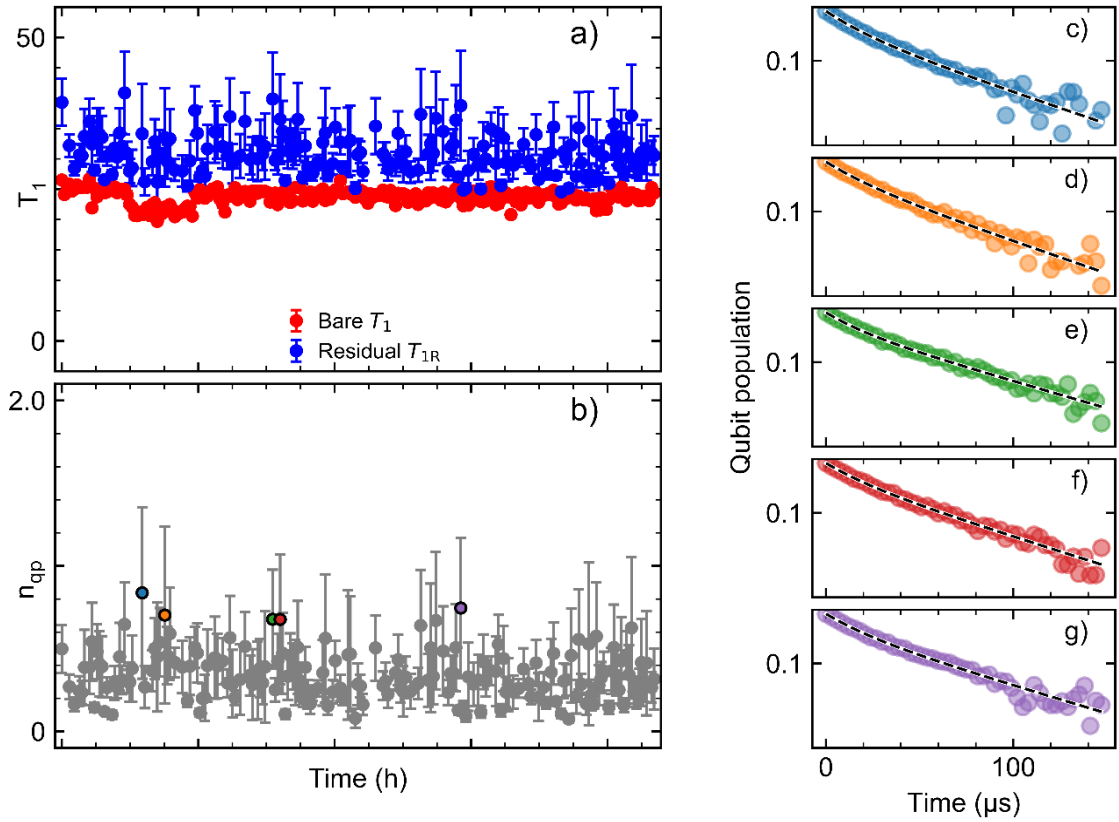

**Supplementary Fig. 5 | Detection of quasiparticle tunneling.** Analysis of $T_1$ measurements performed on qubit A (see Supplementary Fig. 4) via the multiplexer operating at $V_{dd} = 0.7$ V, using the double-exponential model associated with quasiparticle (QP) tunnelling[2] : $y(t) = \exp(n_{qp} (\exp(-t/T_{1qp})-1))*\exp(-t/T_{1R})$, where $y(t)$ is time dependent qubit population, $n_{qp}$ is the quasiparticle density, $T_{1qp}$ is the quasiparticle lifetime, and $T_{1R}$ is the residual relaxation time from other decay channels. The detected quasiparticle density is qualitatively similar to the values reported in Ref. [1], where the measurements were done in the absence of the multiplexer. Therefore, we surmise that the presented cryo-CMOS multiplexer does induce additional Cooper-pair braking mechanisms when placed in proximity and connected to the superconducting transmon qubit.